\documentclass[pra,amsmath,twocolumn]{revtex4}

\usepackage{graphicx}

\begin{document}

\title{Entanglement and Timing-Based Mechanisms in the Coherent 
Control of Scattering Processes}

\author{Michael Spanner}
\author{Paul Brumer}
\affiliation{Chemical Physics Theory Group, Department of Chemistry, 
and Center for Quantum Information and Quantum Control, 
University of Toronto, Toronto, M5S 3H6 Canada}

\date{\today}

\begin{abstract}
The coherent control of scattering processes is considered, with
electron impact dissociation of H$_2^+$ used as an example.
The physical mechanism underlying 
coherently controlled stationary state scattering is exposed by
analyzing a control scenario that relies on previously 
established  entanglement requirements
between the scattering partners.  Specifically, 
initial state entanglement assures
that  all collisions in the scattering volume 
yield the desirable scattering configuration.
Scattering is controlled
by preparing the particular internal state wave function 
that leads to the favored collisional configuration in 
the collision volume.  This insight allows 
coherent control to be extended to the case of time-dependent
scattering. Specifically, we
identify reactive scattering scenarios using 
incident wave packets of translational motion where coherent 
control is operational and initial state entanglement is unnecessary.
Both the stationary and time-dependent scenarios incorporate
extended coherence features, making them physically distinct.
From a theoretical point of view, this
work represents a large step forward in the qualitative understanding
of coherently controlled reactive scattering.  From an 
experimental viewpoint, it offers an  alternative to 
entanglement-based control schemes.  However, both methods
present significant challenges to existing experimental technologies.
\end{abstract}

\maketitle

\section{Introduction}

Coherent control \cite{CCbook} is an approach to controlling
quantum processes where the phase coherence between different
quantum states is explicitly used in order to enhance or suppress
a desired outcome of a given quantum event through interference
effects.  In the case of scattering, 
control over the product cross sections has been shown to be
achievable by creating a coherent superposition of
incident scattering states \cite{LabConditions,degen}.
Unlike the coherent control of unimolecular processes, which has
been established both theoretically and experimentally \cite{CCbook}, 
applications of
coherent control to collisions is only in its infancy.  Hence 
it is important, at this early stage, to clarify and reinforce its
essential principles, as well as to identify and develop 
qualitative pictures and insights applicable to a wide class
of scattering problems, including the all-important case of reactive
scattering.

Previous studies \cite{LabConditions,degen} focusing on crossed beam 
scenarios at fixed total energy
have identified necessary requirements for the coherent 
control of scattering events.   In these stationary state scenarios, incident 
states consisting of coherent superpositions of internal and
translational motions were considered.  In general, due to the 
conservation of center-of-mass momentum and of energy during the collision,
initial state entanglement between the internal and center-of-mass states
of the incident beams was found to be required.  Here entanglement
refers to the non-separability of the initial wave function
when written in terms of the lab frame coordinates of the incident
scattering particles. This requirement
presents a considerable challenge to experimental implementation of
the coherent control of scattering processes, and is 
partially responsible for
the lack of coherent control scattering experiments thus far. 
The need for initial state entanglement in stationary state scenarios 
can be removed if superpositions of degenerate internal
states are used\cite{degen}.  However, reliance upon such 
entanglement-exempt
systems greatly restricts the choice of 
possible situations that can be used in the
coherent control scenario.

Although the formalism leading to conditions for the coherent control
of stationary state 
scattering is clear\cite{LabConditions}, the associated qualitative
insight into the role of initial state entanglement is lacking. In this paper, 
we expose the universal configuration-based mechanism that lies 
at the core of such coherently controlled scattering
\cite{LabConditions}.  Although our results are applicable
to all scattering processes, we focus below on the most
challenging case of reactive scattering.
In particular, by analyzing the generic fixed total energy 
control scenarios  where
entangled interfering pathways are required, it is found that 
entanglement between the two incident beams ensures 
that {\it all} collisions within the scattering volume occur 
at a single favorable configuration of internal 
states.
Further, controlling the relative phases of the 
entangled interfering pathways is shown to shape the  internal 
state wave function that participates in the scattering event.

Having identified this
general mechanism, we extend this thinking to the time-dependent
scattering regime, allowing us to 
introduce an alternate approach to controlled reactive scattering 
that uses non-entangled wave packets of translational motion.
This approach relies on the observation that 
translational wave packets 
have a finite size, so that there is a finite volume of overlap
in both space and time, defined by the scattering beams, wherein
the collisions occur.  If the internal configuration 
does not have time to change as the molecules move across 
the collision region, then a desired configuration can be established
and maintained during the collision time, allowing  
coherent control of reactive scattering
despite the absence of initial state entanglement. 
This method of control relies on the presence of temporal correlations
in the incident state (the collisions need to be precisely 
timed relative to the internal motion) as opposed to initial
state entanglement.

As a result of this work we have exposed the underlying qualitative 
mechanism for coherently controlled scattering and developed it within
both stationary and non-stationary frameworks. Each approach has its own 
challenges in experimental implementation, and each has its individual
benefits. For example, 
the time-dependent wave packet version is perhaps conceptually
and intuitively simpler, while the stationary state
entangled version is not limited to scenarios with 
restricted collision volumes in space and/or time.

In the molecular scattering literature, it is frequently the case
that one uses time-dependent (i.e. wave-packet) methods to 
calculate essentially time-independent scattering properties 
\cite{Zhang}, such as energy-resolved cross sections.
That usage of wave packets is intended solely as a computational
convenience, and the same scattering results could have been
computed using fully time-independent methods.  
We emphasize that, by contrast,  
the stationary and time-dependent scattering control schemes that we
discuss do not simply 
refer to two different computational methodologies used
to calculate the same physical scattering scenario.
Rather, the stationary and time-dependent control scenarios embody 
physically distinct extended coherence properties: one requires
initial state entanglement while the other requires temporal
correlations between the incident wave packet.

Similarly, the time dependent scheme introduced
below does not just consist of simple shaping of a translational 
wavefunction incident on a single state of a target molecule. Such
a scenario would be akin to the previously studied 2PACC 
scenario\cite{2pacc} which we have recently shown\cite{michaelvlado}
does not involve any aspect of quantum
interference, and is hence not coherently controlled collision dynamics.
Rather, as shown below, our time dependent approach
invokes specific interference effects to
affect the product cross sections.

As a working example to illustrate these ideas, we consider
electron impact dissociation of H$_2^+$, where the 
total dissociation cross section and energy spectrum of the 
ejected protons are the observables to be controlled.  
However, it should be stressed that the arguments and conclusions to
follow are completely general and could have been illustrated 
using any scattering example.  Similarly, none of
our results relies on the use of the Born approximation that is invoked below. 
However, the specific example of e-H$_2^+$ 
collisions is focused upon since it is of central 
interest to the emerging field of
strong field attosecond physics \cite{CorkumRecollision},
where laser-induced ionization followed by electron recollision 
with the parent ion lie at the core of many processes.

The paper is organized as follows: Section \ref{Formalism} 
provides a 
reformulation of controlled stationary state scattering which affords 
insight into the origin of conditions for control, and Sect. \ref{Ccreactive}
displays the underlying mechanism for control via computations on 
electron impact dissociation of H$_2^+$. The extension of these qualitative
control insights to the case of time-dependent scattering is discussed in
Sect. \ref{wavepacks}.  Section \ref{summary} provides a summary.

\section{Formalism}
\label{Formalism}
\subsection{General stationary state scattering considerations}
\label{SecGeneralScattering}

Typical crossed molecular beam experiments are well described by
stationary state scattering theory at fixed total energy.
Consider then a few general results from stationary state scattering theory.
Within the $S$-matrix formalism \cite{Child,Taylor} the transition
probability from the initial $|{\bf a}\rangle$ to final 
$|{\bf b}\rangle$ state is
\begin{equation}
	S_{\bf ab} = 
	\left \langle {\bf b} \left |
	e^{-i\int_{-\infty}^{\infty}{\widehat H}dt}
	\right | {\bf a} \right \rangle
	= \delta_{\bf ab} - i T_{\bf ab},
\end{equation}
where ${\widehat H}$ is the scattering Hamiltonian, $T_{\bf ab}$ is the 
transition matrix, and $\delta_{\bf ab}$
represents the unscattered component.  Note that all equations
are written in atomic units $\hbar = m_e = e = 1$. 
The $T$-matrix elements have the form
\begin{equation}\label{EqTMatrix}
	T_{\bf ab} = 2\pi \: \delta(E_{\bf b}-E_{\bf a}) \:
	\delta({\bf K}_{\bf a} - {\bf K}_{\bf b}) \: t_{\bf ab},
\end{equation}
where $E_{\bf a}$ and $E_{\bf b}$ are the energies of the
initial and final states, ${\bf K}_{\bf a}$ and ${\bf K}_{\bf b}$
are the center-of-mass momentum of these states.
The $t_{\bf ab}$ are the ``on-shell" $T$-matrix elements, 
and depend only on the relative momentum, ${\bf k}_{\bf a}$ 
and ${\bf k}_{\bf b}$, and the internal quantum numbers, 
denoted ${\bf a}'$ and ${\bf b}'$, of the incident and 
outgoing states.  Physically, the two $\delta$-functions in
Eq. (\ref{EqTMatrix}) enforce conservation of energy and
of total center-of-mass momentum.  

When the scattering pair is launched 
in a single eigenstate $|{\bf a}\rangle$ of the reactant system,
the differential cross section is
\begin{equation}
    \frac{d\sigma_{{\bf b'}}({\bf a})}{d\Omega} 
	= (2\pi)^4 \Big| (1/|{\bf k}_{{\bf a}}|) t_{\bf ab}  \Big|^2,
\end{equation}
where ${\bf K}_{\bf b} = {\bf K}_{\bf a}$, the magnitude of the
outgoing relative momentum is 
\begin{equation}
	|{\bf k}_{\bf b}| 
	= \sqrt{2(|{\bf k}_{\bf a}|^2/2 + E_{{\bf a}'} - E_{{\bf b}'})},
\end{equation}
and $\Omega$ is the 3D angle of the vector ${\bf k}_{\bf b}$.
The total cross section for a particular reactive 
arrangement channel $n$ is then
found by summing over all internal states belonging to $n$, and
integrating over the solid angle $\Omega$,
\begin{equation}
	\sigma^{(n)}({\bf a}) = \sum_{{\bf b}' \in n} 
	\int \frac{d\sigma_{{\bf b}'}({\bf a})} {d\Omega}
	d\Omega.
\end{equation}
Since the on-shell transition matrix elements naturally 
depend on the initial state $|{\bf a}\rangle$,
one way to enhance/suppress a desired reactive cross section
is to use a single incident state $|{\bf a}\rangle$, and vary it 
until the particular incident state is found that achieves this goal.
This method is termed passive single-state control.  

Experimentally creating a single 
incident scattering eigenstate is not always
an easy task.  Often, internal degrees of freedom of the scattering
pairs are in a thermal distribution of quantum states.
In this situation, cross sections can be controlled to 
some degree by varying, for example, the internal temperature of the 
particles leading to a temperature-dependent cross section
\begin{equation}
	\sigma^{(n)}(T) = \frac{1}{Z}
	\sum e^{-E_{{\bf a}'}/kT} \sigma^{(n)}({\bf a}'),
\end{equation}
where $Z$ is the partition function.
Varying the temperature then changes the distribution of states
that participate in the collision, and hence offers control.
The thermal case can, of course, be identified as a particular
case of scattering from an incoherent 
initial distribution, which in general leads to cross sections
of the form
\begin{equation}\label{EqDistro}
	\sigma^{(n)}(F) = 
	\left(\sum_{\bf a} F({\bf a}) \sigma^{(n)}({\bf a})\right) \Bigg/
	\left(\sum_{\bf a} F({\bf a}) \right),
\end{equation}
where $F({\bf a})$ defines the incoherent distribution of 
incident states.  Since all incoherent distributions 
provide averages over the single-state cross-sections, one
can never achieve greater controllability using incoherent
distributions than that achievable in single-state scattering.
However, the use of coherent superpositions of incident eigenstates, 
instead of incoherent distributions, provides an opportunity 
to do better than the single-state scenario.  

\subsection{Superposition states and coherent control}

The results in Sect. \ref{SecGeneralScattering} may be extended by
considering scattering from an initial superposition state, either
of fixed energy (and hence stationary) or of varying energy content
(and hence non-stationary). 
The full differential cross section for scattering 
from the coherent initial superposition 
\begin{equation}\label{EqPsi0General}
	|\Psi_0\rangle = \sum_{{\bf a}'} \iint dk_{\bf a} dK_{\bf a} \:
	                 C({\bf a}',k_{\bf a},K_{\bf a}) \:
	                 |{\bf a}'\rangle|k_{\bf a}\rangle| K_{\bf a}\rangle
\end{equation} 
is 
\begin{equation}\label{EqDSigma}
    \frac{d\sigma_{{\bf b}'}(\Psi_0)}{d{\bf k}_{\bf b}dK_{\bf b} }
	= (2\pi)^4
    \bigg| \sum_{\bf a'} (1/|k_{\bf a}|) t_{\bf ab} 
	 C({\bf a}',k_{\bf a},K_{\bf a}) \bigg|^2
\end{equation}
where again $K_{\bf a} = K_{\bf b}$,
and the initial relative momenta corresponding to each 
${\bf k}_{\bf b}$ and ${\bf b}'$ is now given by
\begin{equation}\label{EqkaW}
	k_{\bf a} 
	= \sqrt{2(|{\bf k}_{\bf b}|^2/2 + E_{{\bf b}'} - E_{{\bf a}'})}.
\end{equation}
For simplicity, the assumption that all incident 
momenta lie along a single axis (i.e. the initial momenta
of the two particle are antiparallel) was used to arrive at
Eq. (\ref{EqDSigma}).  However, full dimensionality is allowed
in the outgoing states.  Allowing for off-axis incident momenta 
would simply introduce addition integrals over the initial momenta
to Eq. (\ref{EqDSigma}), but the results of the present study  
are otherwise unaffected.
The total cross section for scattering into channel $n$ is
obtained from Eq. (\ref{EqDSigma}) as
\begin{equation}
	\sigma^{(n)}(\Psi_0) = \sum_{{\bf b}' \in n} 
	\iint 
    \frac{d\sigma_{{\bf b}'}(\Psi_0)}{d{\bf k}_{\bf b}dK_{\bf b} }
	d{\bf k}_{\bf b}dK_{\bf b}.
\end{equation}

Within the coherent control approach, 
the relative phases between multiple pathways from the 
initial to final state are used to control the
process.  By controlling the relative phase of the
(complex-valued) $C({\bf a}',k_{\bf a},K_{\bf a})$ in the
initial state Eq. (\ref{EqPsi0General}),
multiple coherent pathways can be created that manifest 
themselves as the sum over ${\bf a}'$ in Eq. (\ref{EqDSigma}).
From the form of the $T$-matrix [Eq. (\ref{EqTMatrix})],
we see that all allowed transitions must conserve center-of-mass
momentum and total energy.  This leads immediately to 
the general requirement for scattering interference: the 
initial states contain {\it on-shell coherence}, that is, in order 
for two initial eigenstates to interfere, they must belong
to a single shell as defined by the $\delta$-functions in the
$T_{\bf ab}$-matrix.  Hence, pathways in Eq.(\ref{EqDSigma})
exhibit on-shell coherence (since $K_{\bf a} = K_{\bf b}$, 
and Eq. (\ref{EqkaW}) is simply a statement of energy conservation)
if the initial superposition Eq. (\ref{EqPsi0General}) contains more
than one state satisfying these on-shell requirements.

In the case of field-free bimolecular scattering governed by 
Eq. (\ref{EqTMatrix}), as will be evidenced
below, superposition states with coherence in the lab frame
translational motion of both incident particles, and
at least one internal mode (minimum 3 degrees 
of freedom in total), is required for interference.  
If static or time-dependent external fields are present
during the collision event then they will modify the on-shell conditions,
through energy and momentum exchange with the particles,
and can lead to less restrictive conditions on the required 
initial-state coherence.  

\subsection{e-H$_2^+$ scattering}

As an example of this formalism we will consider
electron impact dissociation of H$_2^+$.
The Hamiltonian for this scattering problem is given by
\begin{eqnarray}
	{\widehat H} &=& \frac{\widehat{\bf P}^2_1}{2m_p}
	+ \frac{\widehat{\bf P}^2_2}{2m_p} + \frac{\widehat{\bf p}^2_b}{2} 
	+ \frac{\widehat{\bf p}^2}{2}  \\ \nonumber
	& & -\frac{1}{\big|\widehat{\bf R}_1-\widehat{\bf r}_b\big|}
	    -\frac{1}{\big|\widehat{\bf R}_2-\widehat{\bf r}_b\big|}
	    -\frac{1}{\big|\widehat{\bf R}_1-\widehat{\bf r}\big|}
	\\ \nonumber
	& & -\frac{1}{\big|\widehat{\bf R}_2-\widehat{\bf r}\big|}
	    +\frac{1}{\big|\widehat{\bf R}_1-\widehat{\bf R}_2\big|}
	    +\frac{1}{\big|\widehat{\bf r}_b-\widehat{\bf r}\big|},
\end{eqnarray}
where the momentum/position operators with the subscripts '1' and '2'
refer to the two protons, those with the 'b' subscript refer to 
the bound electron, and the remain operators refer to the
incident electron.  In this process, 
an electron with momentum ${\bf p}_i$ is incident on 
an H$_2^+$ molecule of momentum ${\bf P}_i$, which is in an internal 
vibrational and rotational state labeled by ${\bf n}$ = $(\nu,J,m_J)$
with energy $E_{\bf n}$ on the ground electronic state $\Sigma_g$.
All indicated momenta are in the laboratory frame.
During the e-H$_2^+$ collision, the incident electron excites the
bound electron from the bonding to antibonding state,
$\Sigma_g \rightarrow \Sigma_u$, through
the electron-electron Coulomb interaction $\widehat V_{ee}$.
The final state of the scattered particles consists of the 
scattered electron with momentum ${\bf p}_f$ and two protons with
momentum ${\bf P}_1$ and ${\bf P}_2$, one of which carries
the bound electron.  For the purposes of the
scattering calculation, the final state of the two
protons is written in terms of the center-of-mass motion
of the full H$_2^+$ composite ${\bf P}_f = {\bf P}_1 + {\bf P}_2$ 
and the relative motion ${\bf P}_r = ({\bf P}_1 - {\bf P}_2) / 2$ 
corresponding to a continuum state of energy $E = |{\bf P}_r|^2/2\mu$ 
and angular momentum state $L$
on the $\Sigma_u$ surface of H$_2^+$, 
where $\mu = m_p/2$ is the reduced
mass of the molecular ion and $m_p$ is the mass of the proton. 
The $\Sigma_u$ continuum states asymptotically approach the 
free particle momentum states defined by ${\bf P}_1$ and 
${\bf P}_2$ at large distances, but are distorted near the core.

The on-shell transition matrix elements can be evaluated to
first order in the electron-electron interaction (first Born 
approximation) \cite{Kerner,Zare}, 
\begin{equation}
	t_{\bf a'b'} =
	 \left \langle \phi^{(u)}_{E,L},{\bf k}_f \left |
	\widehat V_{ee}
	\right | \phi^{(g)}_{\bf n},{\bf k}_i \right \rangle,
\end{equation}
where the energies are given by
\begin{subequations}
\begin{eqnarray}
	E_{\bf a} &=& \frac{{\bf p}_i^2}{2} + \frac{{\bf P}_i^2}{2m_I}
	+ E_{\bf n}, \\
	E_{\bf b} &=& \frac{{\bf p}_f^2}{2} + \frac{{\bf P}_f^2}{2m_I}
	+ E, 
\end{eqnarray}
\end{subequations}
${\bf K}_j$ and ${\bf k}_j$ ($j=i,f$) are the initial and final 
center-of-mass and relative momenta of the e-H$_2^+$ system
\begin{subequations}\label{EqLabRelative}
\begin{eqnarray}
	{\bf K}_j &=& {\bf p}_j  + {\bf P}_j, \\
	{\bf k}_j &=& \frac{m_I {\bf p}_j - {\bf P}_j}{m_I+1}, 
\end{eqnarray}
\end{subequations}
and $m_I = 2 m_p$ is the mass of the ion.  
The electron-electron interaction matrix elements, 
$\left \langle \phi^{(u)}_{E,L},{\bf k}_f \left |
\widehat V_{ee}
\right | \phi^{(g)}_{\bf n},{\bf k}_i \right \rangle$,
are evaluated in full-dimensionality
using the LCAO approximation
for the bound electron \cite{Zare} as described in  
Appendix A.  We avoid additional approximations
\cite{Kerner,Zare} by using numerical 
wave functions for the radial molecular continuum states,
with the $\Sigma_g$ and $\Sigma_u$ surfaces taken from 
Ref. \cite{BunkinTugov}.

The full differential dissociative cross section is
\begin{widetext}
\begin{eqnarray}
	\frac{d\sigma^{(D)}_L(\Psi_0)}{dEd{\bf k}_fdK}
	&=& (2\pi)^4
    \left| \sum_{\bf n} (1/k_{i 0}) 
	\left \langle \phi^{(u)}_{E,L},{\bf k}_f \left |
    \widehat V_{ee}
    \right | \phi^{(g)}_{\bf n},k_{i0} \right \rangle
    \left\langle \phi^{(g)}_{\bf n},k_{i0} , K|
    \Psi_0 \right \rangle  
	\right|^2,
\end{eqnarray}
\end{widetext}
where 
\begin{equation}
	k_{i 0} = \sqrt{2\left(|{\bf k}_f|^2/2+E
	- E_{\bf n} \right)},
\end{equation}
and the initial momenta are again assumed to be antiparallel.
The cross section $d\sigma^{(D)} /dE$, dependent 
only on the energy of the ejected protons
$E$, is calculated by integrating over the unobserved coordinates
\begin{equation}
	\frac{d\sigma^{(D)}(\Psi_0)}{dE} = 
    \sum_{L=1,3,5...}
	\iint \frac{d\sigma^{(D)}_L(\Psi_0)}{dEd{\bf k}_fdK} d{\bf k}_f dK
\end{equation}
and the total yield is given by 
\begin{equation}
	\sigma^{(D)}(\Psi_0) = \int \frac{d\sigma^{(D)}(\Psi_0)}{dE} dE. 
\end{equation}

\section{Coherent control of reactive scattering}
\label{Ccreactive}
\subsection{Stationary State Scattering: Few-state entangled superpositions}

From the requirement of on-shell coherence,  
the simplest (in terms of number 
of states involved) incident superposition [Eq. (\ref{EqPsi0General})]
that offers on-shell coherence, and hence coherent control,
utilizes two different ${\bf a'}$ states, both with
the same total energy (i.e. internal plus translational energy),  
at the same center-of-mass momentum $K$, i.e. a state of the form:
\begin{equation}\label{EqFirstOnShell}
	|\Psi_{os}\rangle = C_1 |{\bf a}'_1\rangle|k_1\rangle|K\rangle
	               + e^{i\phi} C_2|{\bf a}'_2\rangle|k_2\rangle|K\rangle,
\end{equation}
where $C_1$ and $C_2$ are real. The associated 
differential cross section is then
\begin{eqnarray}
    \frac{d\sigma(\Psi_{os})}{d{\bf b'}dK} &=& (2\pi)^4
    \left| C_1(1/k_1) t_{{\bf a}_1'{\bf b}'} 
	+ e^{i\phi} C_2 (1/k_2) t_{{\bf a}_2'{\bf b}'} \right|^2
	\\ \nonumber
	&=& (2\pi)^4 \bigg[ | C_1(1/k_1) t_{{\bf a}_1'{\bf b}'}|^2
	+ | C_2(1/k_2) t_{{\bf a}_2'{\bf b}'}|^2
	\\ \nonumber
	&+& | C_1(1/k_1) t_{{\bf a}_1'{\bf b}'}|| C_2(1/k_2) t_{{\bf a}_2'{\bf b}'}|
	\cos(\varphi_{{\bf a}_1',{\bf a}_2'}+\phi) \bigg ].
\end{eqnarray}
where $\varphi_{{\bf a}_1',{\bf a}_2'}$ is the phase of 
$ t^*_{{\bf a}_1'{\bf b}'} t_{{\bf a}_2'{\bf b}'}$. 
Varying the relative phase $\phi$ of the two components in 
Eq. (\ref{EqFirstOnShell}) controls 
interferences in the scattered state 
and provides a means of controlling the scattering products.
By tuning $\phi$ one can enhance or suppress the cross 
sections beyond the values attainable with single-state and
incoherent scenarios that utilize the same states, provided that $t_{{\bf a}_1'{\bf b}'}$
and $t_{{\bf a}_2'{\bf b}'}$ are not too dissimilar.

The state $|\Psi_{os}\rangle$ is expressed above in 
center-of-mass and relative coordinates.  In lab
frame coordinates, explicitly for the  example of
e-H$_2^+$ scattering, $|\Psi_{os}\rangle$ becomes
\begin{equation}\label{EqEntangledState}
	|\Psi_{os}\rangle = C_1 |{\bf n}^{(1)}\rangle|P^{(1)}_i\rangle
	                   |p^{(1)}_i\rangle
	             + C_2 e^{i\phi}
	                   |{\bf n}^{(2)}\rangle|P^{(2)}_i\rangle
	                   |p^{(2)}_i\rangle,
\end{equation}
subject to the constraints
\begin{subequations} \label{EqEntangledConditions}
\begin{eqnarray}
	P^{(1)}_i + p^{(1)}_i &=& P^{(2)}_i + p^{(2)}_i \\
	\frac{(P^{(1)}_i)^2}{2m_I} + \frac{(p^{(1)}_i)^2}{2} 
	+ E_{{\bf n}^{(1)}} &=& 
	\frac{(P^{(2)}_i)^2}{2m_I} + \frac{(p^{(2)}_i)^2}{2} 
	+ E_{{\bf n}^{(2)}}.
\end{eqnarray}
\end{subequations}
Note that, in the general case
when the internal state $|{\bf n}^{(1)}\rangle$ and
$|{\bf n}^{(2)}\rangle$ are not degenerate, Eq. (\ref{EqEntangledState})
is an entangled state of translational and internal motion,
as previously discussed\cite{LabConditions}.
For degenerate internal states, Eq. (\ref{EqEntangledConditions})
permit the solution $P^{(1)}_i = P^{(2)}_i$, $p^{(1)}_i = p^{(2)}_i$,
thus removing the initial state entanglement requirement\cite{degen}.

In the 
remainder of this section we analyze this control scenario to 
expose the qualitative mechanism lying at the core 
of the entangled stationary state control, and then use this insight to
introduce a time-dependent non-entangled version of coherently controlled
reactive scattering, described further in the following subsection.

\begin{figure}[t]
    \centering
    \includegraphics[width=\columnwidth]{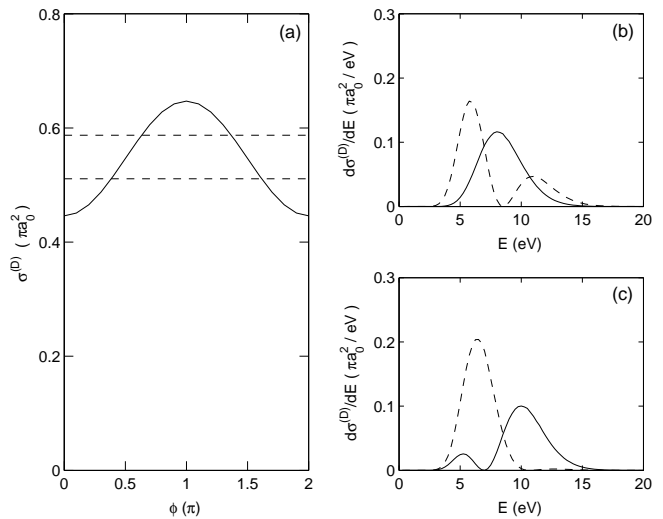}
    \caption{Control of reactive scattering using the entangled
	initial state Eq. (\ref{EqEntangledState}). Panel (a)
	shows the dependence of the dissociation cross section
        $\sigma^{(D)}$ on $\phi$.
	The dashed lines denote $\sigma^{(D)}$ for a single component
	wave function using $\nu=0$ (lower line) and $\nu=1$ (upper line).
	Panel (b) plots the energy spectrum of the proton fragments
	using $\nu=0$ (solid) and $\nu=1$ (dashed).
	Panel (c) shows the analogous spectra when using the
	two-state superposition with $\phi = 0$ (solid) and
	$\phi = \pi$ (dashed).
	} \label{FigNarrow}
\end{figure}

An example of control using the state Eq. (\ref{EqEntangledState})
is shown in Fig. \ref{FigNarrow}. Specifically, 
this example superposes two molecular vibrational 
states with $\nu$ = 0 and 1 and with
energies $E_0$ = -0.0973 au and $E_1$ = -0.0871 au, 
angular momentum $J=m_J=0$, and translational momenta 
$P_i^{(1)}$ = 0 au and
$p_i^{(1)}$ = 4 au
The momenta $P_i^{(2)}$ and $p_i^{(2)}$ are then set  
in accordance with Eq. (\ref{EqEntangledConditions}).
The weights of the two components are set equal, $C_1 = C_2$.
Panel (a) shows the total cross section for dissociation 
as $\phi$ is varied.  For comparison, the two dashed lines show
the cross sections for the $\nu=0$ (lower line) and
$\nu=1$ (upper line) states, considered separately. 
Enhancement or suppression of the
total cross section beyond the incoherent result is clearly 
evident in Fig. \ref{FigNarrow}
when using the entangled coherent superposition 
Eq. (\ref{EqEntangledState}).  
The remaining two panels show the energy resolved cross sections,
$d\sigma^{(D)}/dE$, for the states $\nu=0$ and $\nu=1$ individually 
[panel (a)], and for the superposition states corresponding
to the extrema of $\sigma^{(D)}(\phi)$, namely $\phi$ = 0 and $\pi$
[panel (b)].  A large degree of control over the proton energy 
spectrum is evident.

An important physically transparent qualitative picture of how the 
control arises for the entangled state can be constructed.
To do so, we first note that there is ample evidence \cite{Zhang,kernel}
that the exchange kernels between the reactant and product arrangements
are short range.  As such, the behavior of the wave function when the
reactants are close to one another is particularly relevant to 
the reactive cross sections.  For this reason, we focus below
on the character of the wave function at short range.

\begin{figure}[t]
    \centering
    \includegraphics[width=\columnwidth]{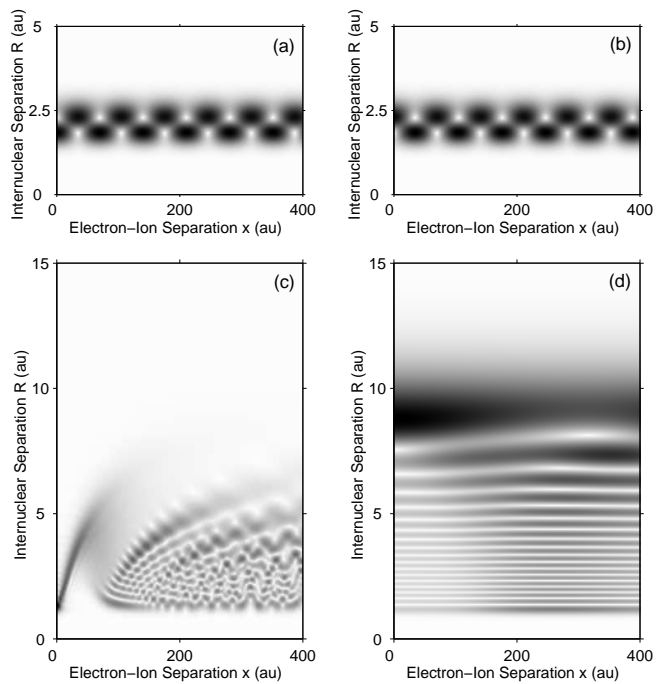}
    \caption{Plots of the probability density of
	relative and internal motion 
	$P(R,x) = |\langle R, x | \Psi\rangle |^2$.
	initial state Eq. (\ref{EqEntangledState}).
	Panels (a) and (b) correspond the case shown in 
	Fig. \ref{FigNarrow} for $\phi = 0$ and $\pi$
	respectively.  Panels (c) and (d) show incident
	states constructed by hand to minimize and maximize the nuclear
	bond length at the moment of collision, and hence
	minimize and maximize $\sigma^{(D)}$.
	} \label{FigMaps}
\end{figure}

In relative and center-of-mass momenta, the incident state is
\begin{equation}\label{EqKk}
	|\Psi_{os}\rangle = \left[ C_1 | \nu^{(1)}\rangle|k_i^{(1)}\rangle
	          + C_2 e^{i\phi} |\nu^{(2)}\rangle|k_i^{(2)}\rangle 
	               \right] |K_i\rangle.
\end{equation}
Since the center-of-mass 
momentum is conserved during the scattering, only the terms in
the square brackets in Eq. (\ref{EqKk}) are relevant for the control 
dynamics.  Figures \ref{FigMaps}a and 
\ref{FigMaps}b show the probability 
 $P(R,x) = |\langle R, x | \Psi_{os}\rangle |^2$, 
of finding $|\Psi_{os}\rangle$ at 
the nuclear bond length $R$ and electron-ion separation $x$ 
(the conjugate of $k_i$).  The internal states and incident 
momenta are the same as those used in Fig.  \ref{FigNarrow};
panel (a) shows results for $\phi$ = 0, while panel (b) uses $\phi = \pi$.  
Note that these plots show the incident states in the 
absence of interparticle interactions; 
they do not include the scattered component. The addition of the 
latter will change our argument quantitatively but not qualitatively.
As noted above, of particular interest is the character of
the internal state near the collision region, $x \approx 0$.
Since $|\Psi_{os}\rangle$ 
is a time-independent wave function the probabilities plotted
in Figs. \ref{FigMaps}a and \ref{FigMaps}b reflect the complete
incident dynamics of the scattering pair.
Panels (a) and (b) show that the 
internal vibrational wave function near $x = 0$,
in fact for all $x$, is controlled by $\phi$.  
Further, from Eq. (\ref{EqKk}), it is seen that the
wave function at $x = 0$ is
\begin{equation}
	\langle R,x=0|\Psi_{os}\rangle = 
	\left[ C_1 \langle R| \nu^{(1)}\rangle
	     + C_2 e^{i\phi} \langle R |\nu^{(2)}\rangle \right], 
\end{equation}
which, by varying $\phi$, can be shaped into structures 
not accessible using individual incident eigenstates.  Since the 
cross section is strongly dependent on the internal wave function,
controlling $\phi$ then allows one to control $\sigma^{(D)}$
and $d\sigma^{(D)}/dE$ by manipulating the particular internal 
configuration that participates in the collision.  This is the 
qualitative mechanism lying 
at the core of coherently controlled reactive scattering 
using the entangled superposition state Eq. (\ref{EqEntangledState}):
The initial superposition state defines the structure
of the internal state wave function at the point of collision
and can therefore be used to tune this structure in order to 
optimize the reactive cross section.

In standard scattering theory one often considers scattering
of incident time-independent eigenstates, and hence this observation
may seem trivial.  This is not the case.
Had a non-entangled wave function of the form 
\begin{eqnarray}
	|\Psi^{(ne)}_{os}\rangle &=& [ C_1 |{\bf n}^{(1)}\rangle 
	               + C_2 e^{i\phi_a} |{\bf n}^{(2)}\rangle ] \\ \nonumber
	               &\times&
	               [ C_3 |P^{(1)}_i\rangle
	               + C_4 e^{i\phi_b} |P^{(2)}_i\rangle ] \\ \nonumber
	               &\times&
	               [ C_5 |p^{(1)}_i\rangle
	               + C_6 e^{i\phi_c} |p^{(2)}_i\rangle ] \\ \nonumber
\end{eqnarray}
been used, the resultant incident state would contain 8 terms, 
only two of which are at the same total energy and center-of-mass
momentum, namely the components in Eq. (\ref{EqEntangledState}).  
From an energy domain perspective,
this means that only two of the total eight terms exhibit on-shell
coherence, and hence only two of the eight terms can interfere.  The
remaining six terms contribute incoherently to the cross sections.
From a time domain perspective, only the superposition of the two 
on-shell components lead to a time-independent wave function.  All
other components, since they are at different energies,  
accumulate a time-dependent phase relative to the on-shell 
superposition.  With respect to the internal configuration
near the collision region, this means that the off-shell components 
will alternate between constructive and destructive interferences 
at different times, introducing a time-average over all possible 
phases relative to the on-shell component, and hence an average
over internal configurations allowed by the populated
internal states at the collision point.
These time-varying interferences reduce the off-shell 
contribution to an effective 
incoherent contribution.  The entanglement in the few-state 
superposition Eq. (\ref{EqEntangledState}) then plays a
crucial role in limiting the particular internal state wave functions
that participate in the scattering events.

Further investigation of the e-H$_2^+$ cross section strengthens
the configuration-based qualitative mechanism
described above.  In particular, it is known that the 
electron impact dissociation cross section of H$_2^+$ is a 
strongly dependent monotonic function of bond length $R$;
as $\langle R \rangle$ increases, so does $\sigma^{(D)}$ \cite{Peek}. 
Comparing the $x=0$ structure in Figs. \ref{FigMaps}a and
\ref{FigMaps}b shows that the control of $\sigma^{(D)}$ follows 
the expected $\langle R \rangle$-dependence; the 
$\phi=\pi$ case has both the larger $\langle R \rangle$ and 
$\sigma^{(D)}$.  

Using this insight, a clear extension of the two-state
entangled coherent control scenario to a multi-state entangled 
scenario, in order
to achieve better enhancement/suppression than that of Fig. \ref{FigNarrow},
can be constructed.  Since control here
is directly linked to the average bond length at
$x=0$, the superpositions that maximize/minimize the 
scattering cross sections are the same superposition that maximize/minimize 
$\langle R \rangle$ at $x=0$.  For example, two superpositions 
(constructed by hand) that approximately achieve these goals are
\begin{equation}\label{EqMax}
	|\Psi_{\rm max}\rangle = N_{\rm max} \left[ \sum_\nu (-1)^\nu
	                         e^{-\left(\frac{\nu-18}{1.8} \right)^2 }
	                         |\nu\rangle| k_\nu\rangle \right]
	|K\rangle
\end{equation}
and
\begin{equation}\label{EqMin}
	|\Psi_{\rm min}\rangle = N_{\rm min} \left[ \sum_\nu 
	                         e^{-\left(\frac{\nu-7}{6} \right)^2 }
	                         |\nu\rangle|k_\nu\rangle \right]
	|K\rangle
\end{equation}
where $N_{\rm max}$ and $N_{\rm min}$ are normalization constants,
$k_0$ = 4 au, and the remaining $k_\nu$ are set to
ensure that all components satisfy the on-shell requirement, having
the same total energy and
center-of-mass momentum.
The summation runs over all the vibrational states $|\nu\rangle$. 
The corresponding $P(R,x)$ are shown in Figs. \ref{FigMaps}c and
\ref{FigMaps}d.  For these states, the values of $\langle R \rangle$
in the collision region are 
$\langle \Psi_{\rm max} |R| \Psi_{\rm max} \rangle_{x=0} = 8.75$ au
and 
$\langle \Psi_{\rm min} |R| \Psi_{\rm min} \rangle_{x=0} = 1.32$ au, 
while the largest and smallest values possible
using an incoherent initial state would correspond to those
of the highest and lowest vibrational states of $H_2^+$, i.e.,
$\langle \nu=18 |R| \nu=18 \rangle_{x=0} = 8.12$ au
and
$\langle \nu=0 |R| \nu = 0 \rangle_{x=0} = 2.05$ au
The coherent superpositions $|\Psi_{\rm max}\rangle$ and
$|\Psi_{\rm min}\rangle$ access values of 
$\langle R \rangle$ at $x= 0$ beyond those accessible to 
any incoherent mixture of states, and hence provide more control
over $\sigma^{(D)}$ than any incoherent scenario.

Note that although in the case of e-H$_2^+$ scattering there
is a clear classical explanation of the nature of the configuration 
that enhances control, this is need not be the case in general. Rather,
given the Hamiltonian $\widehat H$ and the associated scattering matrix
$S_{\bf ab}$, there is a well defined scattering configuration that
maximizes control. When the on-shell (and hence entanglement)
requirements are met for the initial superposition [Eq. (\ref{EqPsi0General})]
then the varying the coefficients
$C({\bf a}',k_{\bf a},K_{\bf a})$ in 
Eq. (\ref{EqPsi0General}) alters the stationary spatial
configuration and hence
alters the cross sections. The optimal choice 
of the $C({\bf a}',k_{\bf a},K_{\bf a})$
then corresponds to the superposition that comes closest to the
optimal stationary state configuration for scattering.
Whether the optimal configuration is easily	      
understood classically, however, depends upon the system under
consideration.

\subsection{Non-entangled wave-packet superpositions: Time dependent scattering}
\label{wavepacks}
\begin{figure}[t]
    \centering
    \includegraphics[width=\columnwidth]{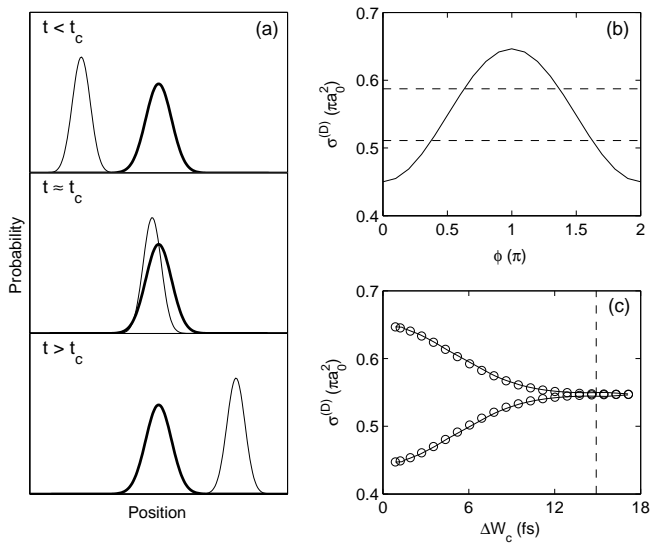}
    \caption{Control of reactive scattering using the wave packets
	of translational motion.  Panel (a) shows schematically 
	the wave packet scenario: controlling the internal state
	of the molecular ion during the finite duration of the
	collision allows for control. $t_c$ denotes the time of the
        collision.  Panel (b) shows the degree
	of control using the initial Eq. (\ref{EqNonEntangled}).
	See the text for the specific parameters used.  Panel (c)
	demonstrates the loss of control as the duration of the
	collision between the two wave packet approaches the
	timescale of the internal state motion.  The dashed
	line denotes the vibrational period of H$^+_2$, 
	$\tau_{vib} = 14.9$ fs.
	} \label{FigWavePackets}
\end{figure}

Having exposed the qualitative mechanism that underlies
control resulting from stationary state, and hence entangled, initial 
superposition states, additional approaches to coherent control
of reactive scattering that do not rely on initial state entanglement can be 
identified.  For example, there is an alternative route to ensuring 
that collisions between two particles occur at a fixed phase of the
internal-state motion: design time-dependent superposition 
states (i.e. Eq. (\ref{EqPsi0General}) with non-degenerate states),
specifically wave packets
of translational motion plus superpositions of internal states,
in order to localize the collision
partners in space, and thereby to restrict the duration of the
collision between the wave packet to less than the internal state 
motion (see Fig. \ref{FigWavePackets}a). Quantum interference manifests
in this case due to the numerous energetically degenerate sets of
states that occur due to the energy widths of the two incident wave
packets. 

The initial state used to illustrate the wave packet
scenario is 
\begin{equation}\label{EqNonEntangled}
	|\Psi_W(t=0)\rangle = |\psi_\nu\rangle |\psi_{p_i}\rangle
	                 |\psi_{P_i}\rangle,
\end{equation}
where 
\begin{subequations}
\begin{equation} \label{EqVibStates}
	|\psi_\nu\rangle = \left[ |\nu=0\rangle + e^{i\phi}|\nu=1\rangle \right]
	/ \sqrt{2},
\end{equation}
\begin{equation}
	|\psi_{p_i}\rangle = \int dp_i 
	 (\Delta_{p_i} \sqrt{\pi} )^{-\frac{1}{2}} 
	e^{ -\frac{1}{2}\left( \frac{p_i-p_{i0}}{\Delta_{p_i}} \right)^2 }
	| p_i \rangle,
\end{equation}
and
\begin{equation}
	|\psi_{P_i}\rangle = \int dP_i 
	(\Delta_{P_i} \sqrt{\pi} )^{-\frac{1}{2}} 
	e^{-\frac{1}{2}\left( \frac{P_i-P_{i0}}{\Delta_{P_i}} \right)^2}
	| P_i \rangle,
\end{equation}
\end{subequations}
as depicted in schematically in
Fig. \ref{FigWavePackets}a.  Note that this is not simply wavepacket scattering
off of a single vibrational state. Rather, scattering is off an internal
superposition of states, necessary to incorporate interference and achieve
coherent control.

To get an idea of how long 
the collision between the two 
wave packets lasts, the following measure, here called
the time-dependent collision probability $W_c(t)$, is used 
\begin{equation}\label{EqPct}
	W_c(t) = N \left|\left\langle\Psi(t)\left|
	\delta({\bf x}-{\bf y})\right|\Psi(t)\right\rangle\right|^2
\end{equation}
where ${\bf x}$ and ${\bf y}$ are the positions of the
electron and ion respectively, $|\Psi(t)\rangle$ is the 
incident wave function (i.e. scattered components are not included)
and $N$ is a normalization constant such that $\int W_c(t) dt = 1$.
This quantity gives the probability of finding the electron 
and the ion at the same position in space, and hence reflects
the probability of a collision occurring at time $t$.  
In the continuous beam scenario, as considered in the previous section,
$W_c(t)$ is a time-independent
constant.  However for wave packet collisions, $W_c(t)$ will 
be a localized Gaussian-like function indicating that, 
in this case, collisions only occur during a select time window 
where the two colliding wave packets overlap in space.
The duration of the wave packet collision is then defined as
twice the standard deviation of $W_c(t)$
\begin{equation}
	\Delta W_c = 2 \sqrt{ \langle t^2 \rangle_{W_c} 
	                          - \langle t \rangle^2_{W_c} },
\end{equation}
where $\langle \cdots \rangle_{W_c}$ indicates an average value where 
$W_c(t)$ is used as the distribution function.
  
In order to compare the wave packet scenario to the entangled state
scenario, we set the mean incident momenta of 
Eq. (\ref{EqEntangledState}) to $P_{i0} = 0$ and $p_{i0} = 4$ au, 
the same as used above.  Using then momentum widths of
$\Delta_{P_i} = 1$ au and $\Delta_{p_i} = 0.01$ au
gives a scattering scenario where $\Delta W_c = 0.87$ fs.
Since $\Delta W_c$ in this case is much smaller than the vibrational
period of H$^+_2$ ($\tau_{vib} = 14.9$ fs), we expect that
control is possible using these parameters.  Figure 
\ref{FigWavePackets}b plots the 
corresponding reactive cross section as $\phi$ 
is varied.  Control is indeed present.
Further, upon comparing Figs. \ref{FigWavePackets}b and 
\ref{FigNarrow}a, one sees that the control using the non-entangled
state Eq. (\ref{EqNonEntangled}) gives the same degree of control
as the entangled state Eq. (\ref{EqEntangledState}) confirming
that the wave packet scenario offers an alternate 
and equivalent (in terms of the controllability of the total 
cross section) means of coherent control of reactive scattering, 
at least for reactive scattering absent of any resonances related
to the relative momentum $k_i$.  This latter point will be
revisited in the following section and is at the root of possible
advantages of using the entangled scenario.

The remaining panel of Fig. \ref{FigWavePackets} plots the 
minimum and maximum values of the cross section, which
correspond to $\phi$ = 0 and $\pi$, as the duration of the
wave packet collision $\Delta W_c$ is increased.  Two different 
methods of increasing $\Delta W_c$ were explored.  For the first 
method, $\Delta p_i$ is gradually made smaller and smaller, which 
causes the spatial size of the electron wave packet to increase 
and thereby increases $\Delta W_c$.  The control results
using this technique, plotted against $\Delta W_c$,  
are shown in Fig. \ref{FigWavePackets}c
using the solid curves.  The second method offsets the spatial
and temporal focus of the electron wave packet relative 
to the ion wave packet
\begin{eqnarray}
	|\psi_{{\bf p}_i}\rangle &=& \int dp_i 
	 (\Delta_{p_i} \sqrt{\pi} )^{-\frac{1}{2}} 
	e^{ -\frac{1}{2}\left( \frac{p_i-p_{i0}}{\Delta_{p_i}} \right)^2 }
	\\ \nonumber
	& \times & e^{ i p_i x_d } \:
	e^{-i (p_i^2/2) \tau_d }
	| p_i \rangle
\end{eqnarray}
where $x_d \equiv p_{i0} \tau_d$, and $\tau_d$ controls the focusing offset.
The control results using this method are plotted in Fig.
\ref{FigWavePackets}c with circles.  In both cases, the control
goes to zero as the $\Delta W_c$ approaches the vibrational 
period $\tau_{vib}$.  These calculations most clearly demonstrate
the configuration-based mechanism underlying the control; 
the controllability decreases as more internal configurations participate
in the collisional events.

\subsection{Entangled or Wave Packets?}

Both the stationary state and non-stationary state control schemes
present technological challenges. The wave 
packet case requires, experimentally, a collision over very short times, 
and therefore must be run in either a pulsed mode, or with very 
tightly focused molecular beams with small overlap region such 
that the molecules move through the collision region
in a time smaller than the timescale characteristic of the 
dynamics of the internal superposition. 
The stationary state case, on the other hand, requires initial state entanglement,
but can be used in a 
continuous beam regime and/or with arbitrarily large 
spatial region of overlap of the two beams.  Hence, the entangled
version offers arbitrarily large space-time scattering volumes
while spatial restrictions exist for the wave packet version.  This
implies that, although the cross sections may be controlled to 
the same degree, the entangled version will always permit
larger total yields since it came be used in conjunction with
arbitrarily large volumes and incident fluxes. 

A second advantage of the entangled scenario relates 
to the possibility of scattering resonances.
The reactive cross section for e-H$^+_2$ scattering is a rather
smooth function of the incident relative momentum $k_i$.  Hence,
our sample cases thus far have implicitly considered
the control of reactive scattering in the absence of sharp
resonances related to the incident kinetic energy.  However, 
reactive scattering problems of chemical interest may
exhibit such resonances (e.g. Feshbach resonances).  The pulsed
wave packet scenario requires a broad superposition of 
incident momenta in order to obtain spatial localization 
and short-time overlaps of the two incident wave functions, while
the entangled case can use as few as two incident momenta.
It may very well be the case that narrow resonances can be
resolved/exploited in the entangled case
\cite{vlado}, while averaging over the 
broad momentum bandwidth in the wave packet case washes
out the narrow resonance features, rendering them unusable 
in this latter scenario.

Note that both of these aspects of initial state entanglement, the possibility
to efficiently exploit narrow resonances and access arbitrarily
large scattering volumes, represent
non-classical aspects of controlled reactive scattering 
accessible via entanglement.

\section{Summary}
\label{summary}

Previous work on stationary state coherent control,
which provides a useful description of crossed beam experiments,
was shown to require initial state entanglement between the 
incident translational and the internal states of the scattering 
partners\cite{LabConditions}.  This paper has provided 
physical insight into the role of this requirement,
generating an extension to coherent control via time-dependent
wave packet scattering. 
Specifically, the initial state entanglement was shown to
assure that  {\it all} collisions occur at a fixed configuration
of the internal state motion.  This qualitative insight then allows 
for the introduction of a coherent control scenario in time-dependent
scattering.  Control
in the latter regime is possible if the duration of the
wave packet collision is much smaller than the characteristic
timescales of motion of the superposition of internal states.

The mechanism of fixed-configuration scattering 
underlying coherent control of reactive scattering
can be extended to all scattering scenarios.  A sample extension
to scattering off surfaces is provided elsewhere\cite{michaelvlado}.

Some extensions of this approach are worth noting. First,
scattering studies carried out 
on loosely bound van der Waals complexes \cite{Wittig}
are distantly related to the wave packet control scenario.
In these
studies, a CO$_2$ $\cdot$ HBr complex was used as an oriented 
precursor to study the CO$_2$ + H reaction, where  photodissociation
of HBr launched the H toward the CO$_2$.  These experiments
are examples of the selection of 
well defined angular wave packets of the scattering
partners.  However, for active control of 
reactive scattering in these systems, one would also need to
tune the particular angular wave packets that participate, as 
opposed to selecting the single orientation defined by the
initial van der Waals complex.  This could perhaps be accomplished by
selectively exciting rotational states of the CO$_2$ and/or HBr
prior to the initiation of the reaction.

Second, although not explored in this paper, both the entangled
and wave packet scenarios explored herein 
could be used instead to completely characterize the
scattering matrix $S_{\bf ab}$ \textit{with phases}, in analogy with
methods of Quantum Process Tomography \cite{QPT}.  In short, 
having selective control over the input internal state,
in addition to the usual control over the input translational momenta,
would allow one to measure enough projections 
of the scattering operators, through measurements of the
scattering cross sections for different internal superpositions, 
to allow for an accurate reconstruction of the complex $S$-matrix.

\section{Acknowledgments}
This work was support through a Discovery Grant from the National 
Science and Engineering Research Council (NSERC) of Canada.

\appendix

\section{Electron Impact Matrix Elements}

For both the strong field and field-free scenario, the 
transition matrix elements 
$\langle \Psi_{out} | \widehat V_{ee}| \Psi_{in}\rangle$
are required.  In calculating these values, we use the
LCAO approximation
for the bound electron\cite{Kerner,Zare}.  Specifically, using
a basis of definite energy and angular momentum for the
final state of the internuclear coordinate ${\bf R}$ 
and restricting the initial H$_2^+$ to zero 
angular momentum gives
\begin{equation}\label{EqVee}
	\langle \phi^{(u)}_{E,L},{\bf k}_f|
	\widehat V_{ee}
	|\phi^{(g)}_{\bf n},{\bf k}_i\rangle  =
	 i^L \sqrt{(2L+1)} 
	 {\cal R}(L,\nu,E,\tilde k),
\end{equation}
where $\tilde {\bf k} \equiv {\bf k}_f-{\bf k}_i$,  
$\tilde k = |\tilde {\bf k}|$, and
\begin{eqnarray}
	{\cal R}(L,\nu,E,\Delta k) = 
	\frac{16}{\pi^2} \frac{1}{{\tilde k}^2 [4+{\tilde k}^2]^2}
	\times \nonumber \\
	\int N^{(+)}(R) N^{(-)}(R)  \chi_{E,J}(R)j_L({\tilde k}R/2)
	\chi_\nu(R) dR.
\end{eqnarray}
The z-axis of the angular states lies along $\Delta {\bf k}$, and 
only the $m_L=0$ sub-levels along this axis have non-zero amplitude.
The $N^{(\pm)}(R)$ are normalization factors arising from 
the LCAO wave functions of the bound electron and are given by
\begin{equation}
	N^{(\pm)}(R) = \left[2\pm2e^{-R}(1+R+R^2/3)\right]^{-1/2}.
\end{equation}
The $\chi_\nu(R)$ are the bound radial eigenstates on the $\Sigma_g$
surface 
\begin{equation}
	\left[ -\frac{1}{2\mu}\frac{\partial^2}{\partial R^2} 
	+ V_{\Sigma_g}(R)  - E_\nu \right] \chi_\nu(R) = 0
\end{equation}
while the $\chi_{E,L}(R)$ are the continuum radial eigenstates
on $\Sigma_u$
\begin{equation}
	\left[ -\frac{1}{2\mu}\frac{\partial^2}{\partial R^2} 
	+ V_{\Sigma_u}(R) + \frac{L(L+1)}{2 \mu R^2} - E\right] \chi_{E,L}(R) = 0.
\end{equation}
The $V_{\Sigma_g}(R)$ and $V_{\Sigma_u}(R)$ surfaces are taken
from Bunkin and Tugov \cite{BunkinTugov}.  $V_{\Sigma_g}(R)$ is
written in Morse oscillator form, and hence analytical
Morse oscillator states are used for 
$\chi_\nu(R)$, while the $\chi_{E,L}(R)$
are integrated numerically and normalized such that
(see Child, Appendix A \cite{Child})
\begin{equation}
	\int_0^\infty \chi^*_{E',L}(R) \chi_{E,L}(R) dR 
	= \delta(E-E').
\end{equation}


\end{document}